# A single-crystal source of path-polarization entangled photons at non-degenerate wavelengths

S. Sauge[*], M. Swillo, M. Tengner, A. Karlsson

[1]*Department of Microelectronics and Information Technology*
*Royal Institute of Technology, KTH, Electrum 229, SE-16440 Kista, Sweden*
[*]sauge@kth.se

**Abstract:** We demonstrate a bright, narrowband, compact, quasi-phase-matched single-crystal source generating path-polarization-entangled photon pairs at 810 nm and 1550 nm at a maximum rate of $3 \times 10^6$ s$^{-1}$ THz$^{-1}$ mW$^{-1}$ after coupling to single-mode fiber, and with two-photon interference visibility above 90%. While the source can already be used to implement quantum communication protocols such as quantum key distribution, this work is also instrumental for narrowband applications such as entanglement transfer from photonic to atomic qubits, or entanglement of photons from independent sources.

**1. Introduction**

Two-photon entangled states manifest one of the most striking phenomena at the heart of quantum mechanics. They have been used to demonstrate the violation of Bell inequalities and thus exclude local realism from the quantum theory [1,2]. They now constitute an integral resource for the implementation of quantum information protocols, such as quantum key distribution [3], quantum teleportation [4,5], or all-optical quantum computing [6-8]. They are also instrumental for precision measurements beyond the standard quantum limit [9].

Different approaches can be considered to create entangled photon pairs. Promising networkable sources under development generate photons directly in single-mode fibers [10, 11] or photonic crystal fibers [12] by means of four-wave mixing. Other emerging sources use semiconductor components with the prospect of integration on optical chips. Progresses have thus been reported with sources based on biexciton cascade of a quantum dot [13] or spontaneous parametric down-conversion (SPDC) and phase matching in semiconductor waveguides [14] or non-linear photonic crystals [15]. While those sources have the potential of increasing the generation efficiency, the most practical and spectrally bright sources to date are based on SPDC in non-linear dielectric quasi-phase matched (QPM) crystals [16-18] and waveguides [19,20], for which the periodic modulation of the $\chi^{(2)}$ nonlinearity allows collinear, wavelength-tunable emission of the photon pairs.

SPDC can be used under type-I (co-polarized outputs) or type-II (orthogonally polarized outputs) phase matching. Bi-directionnal pumping of one single crystal provides quasi-spectral and spatial indistinguishability between counter-propagating pairs, and given that temporal indistinguishability is also ensured between them, photons can be entangled in polarization, resulting in states of the form $|HH\rangle + e^{i\varphi}|VV\rangle$ (type-I) or $|HV\rangle + e^{i\varphi}|VH\rangle$ (type-II), where H and V denote horizontal and vertical polarizations, respectively, and $\varphi$ is the output phase factor, which can be tuned to generate any of the four maximally entangled Bell states. Incidentally, the existence of a fixed phase relation between the two terms relies on the impossibility of deducing the polarization of a photon from any of its other properties after recombination of the two interfering paths.

In the case of type-II SPDC, inserting the non-linear crystal inside a Sagnac interferometer provides an intrinsically phase-stable source [21-24], because both counter-propagating beams travel the same loop structure, so that any change of path length in the loop is experienced by both beams, resulting in stable interferences. While it was claimed that such a set-up could work with photon pairs created at non-degenerate wavelengths, experimental realizations so far have been limited to operation near degeneracy [21-24].

In this paper, we present a type-I QPM single-crystal source of entangled photons operating at non-degenerate wavelengths. Photons are entangled in path and polarization at 810 and 1550 nm, respectively. The path-entanglement configuration at 810 nm leads to a very compact set-up, while the use of polarization coding at 1550 nm with only 0.8 nm bandwidth makes the fiber-coupled source potentially suitable for long-distance quantum communication in field experiments, in view of the possibility to achieve real-time polarization control of each flying qubit, as proposed recently [25], by means of two narrow-spaced reference signals multiplexed in wavelength with the narrowband quantum channel. The single-crystal source has a spectral brightness of $3 \times 10^6$ $s^{-1}$ $THz^{-1}$ $mW^{-1}$ after coupling to single-mode fiber, about ten times larger than the one previously reported for operation at non-degeneracy [17], with similar two-photon interference visibility above 90%, and a reduced number of components. While the set-up is no more intrinsically phase-stable, the source configuration nevertheless provides the means to lock the phase by active stabilization of a Mach-Zehnder interferometer traveled by the pump light, as will be shown. This is a further improvement in contrast to [17]. Providing a narrower bandwidth of tens of pm, which could be achieved in the same configuration with a cavity-inserted waveguide, the source could also be used for applications such as entanglement transfer from photonic to narrowband atomic qubits at around 800 nm [26], or entanglement of (temporally indistinguishable thus narrowband) photons from two independent sources [27], as required for the realization of a quantum repeater.

## 2. The source

The source is drawn in Fig. 1. We use a 50 mm-long type-I bulk crystal made of periodically poled (PP) lithium niobate PPLN:MgO with a grating period of 7.5 μm (HC-Photonics). The crystal is pumped by two focused counter-propagating beams driven by a diode laser operating at a wavelength of 532 nm (Cobolt). The source has collinear emission at the non-degenerate wavelengths of $\lambda_s$ = 810 nm for the signal and $\lambda_i$ = 1550 nm for the idler, allowing efficient single-photon detection and low attenuation in fibers at telecom wavelength, respectively (wavelength tuning can be achieved by varying the temperature of the brass oven heating the crystal at around 100 °C with 0.1°C stability).

The pump is focused into the crystal by an achromatic lens (f= 150 mm) and split in two orthogonal beams by a polarizing beam-splitter (PBS). The intensity of the pump along the two arms is balanced by means of a half-wave plate (HWP) set before the PBS. Since the crystal only down-converts pump light having a polarization of the electromagnetic field set along the crystal's optical axis, a second HWP is used along one of the two pump arms in order to undo the polarization rotation induced by the PBS. The signal and idler are generated with vertical (V) polarization. For the pump beam, we estimated a waist at the focus point of $2w_0$ = 250 μm, giving a Rayleigh range $z_0 = \pi w_0^2/\lambda$ of about 90 mm in PPLN:MgO, hence a $z_0/L$ ratio of 1.8.

The two counter-propagating idler beams generated at 1550 nm are recombined at the PBS splitting the pump, after rotation of the polarization in one of the two arms by the same HWP used to adjust the polarization of the pump beam. Both PBS and HWP are optimized for idler wavelength only because the resulting intensity imbalance between the vertically-polarized components of the two pump beams down-converting in the crystal can be compensated by adjusting the HWP at 532 nm, and by increasing the pump power. After the PBS, the idler is separated from pump by means of a dichroic mirror, filtered from residual pump light by an isolator, coupled to 100 meters of single-mode fiber after which one analyzes the polarization state of the qubit with a PBS and a motorized HWP used for the purpose of recording two-photon visibility curves. A polarization controller is used to align the polarization at the output of the fiber with respect to the analyzer. The bandwidth of the 1550 nm photon after coupling to single mode fiber, as measured by a spectrograph at full pump power without conditional gating, is 0.8 nm FWHM, compatible with 100 GHz wavelength multiplexing.

The two signal beams generated at 810 nm are separated from the pump beams by dichroic mirrors positioned near the exit of the QPM crystal, and the correlations shared with idler photons are mapped onto the four output ports of three non-polarizing beam-splitters (BS) before coupling to one of four avalanche photo-detectors (Si APD) labeled H, V, D and A, in relation to the horizontal (H), vertical (V), diagonal (D) and anti-diagonal (A) polarization states of the idler qubits measured at 1550 nm in the two conjugate basis H/V and D/A, which in this case are selected by rotation of the HWP, see Fig. 1.

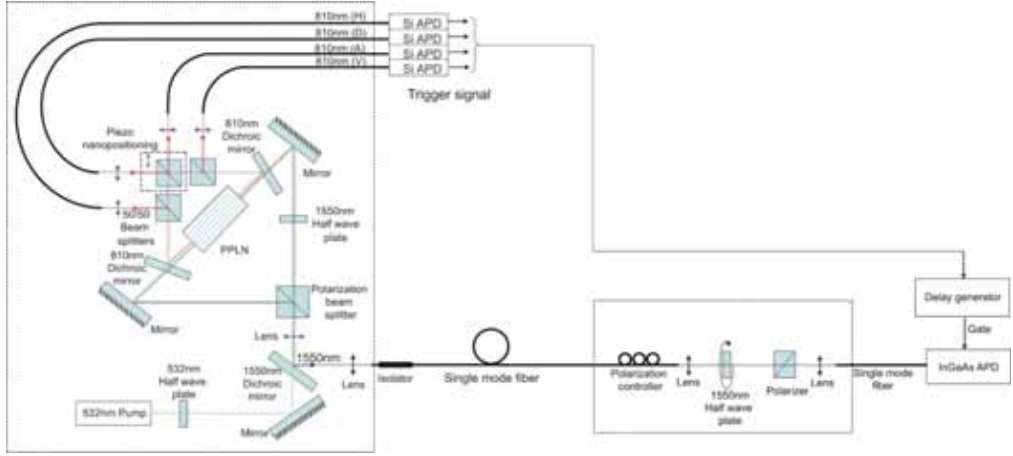

Fig.1. Single crystal source of entangled photons at 810 and 1550 nm.

If APD "H" or "V" clicks when detecting a photon at 810 nm, the transmitted qubit at 1550 nm must have H or V polarization, respectively, since the path taken by the vertically polarized signal photon at 810 nm determines non-ambiguously the polarization acquired by the corresponding idler photon as it reaches the PBS. In this case, the bi-photon state can be written

$$|\Phi^\varphi\rangle = 1/\sqrt{2}\ (|H(\omega_s)\ H(\omega_i)\rangle + e^{i\varphi}\ |V(\omega_s)\ V(\omega_i)\rangle), \qquad (1)$$

where $|H(\omega_s)\rangle$ and $|V(\omega_s)\rangle$ denote the states of the signal photons exiting the output ports of the beam splitters leading to detectors "H" and "V", respectively, $|H(\omega_i)\rangle$ and $|V(\omega_i)\rangle$ denote the polarization states of the idler, and $\varphi$ is a phase factor. $\omega_i + \omega_s = \omega_p$, where $\omega_p$, $\omega_s$ and $\omega_i$ are the pump, signal and idler frequencies, respectively,

If the signal beams at 810 nm interfere at the third 50-50 beam splitter, then depending on the value of $\varphi$ coincidences or anti-coincidences will be observed when measuring idler qubits in the 45° (D/A) rotated polarization basis where they can interfere. To illustrate this, let $|l_1(\omega_s)\rangle$ and $|l_2(\omega_s)\rangle$ refer to the states of the signal photon in the arms of the interferometer set between the crystal and the third BS. If $|D(\omega_s)\rangle$ and $|A(\omega_s)\rangle$ denote the state of the signal photon exiting the output ports of the BS leading to detectors "D" and "A", respectively, then up to a relative phase factor already included in $\varphi$, $|l_1(\omega_s)\rangle$ and $|l_2(\omega_s)\rangle$ can be written

$|l_1(\omega_s)\rangle = r |D(\omega_s)\rangle + t |A(\omega_s)\rangle$ and $|l_2(\omega_s)\rangle = t |D(\omega_s)\rangle + r |A(\omega_s)\rangle$, where $r$ and $t$ denote the reflection and transmission amplitudes of the beam splitter, respectively. Assuming a 50-50 symmetric beam splitter, $t$ leads $r$ in phase by $\pi/2$ and $t = ir = i/\sqrt{2}$, with $i^2 = -1$ [28]. For the idler photons, the $|H(\omega_i)\rangle$ and $|V(\omega_i)\rangle$ polarization states can interfere in the D/A basis, resulting in either of the states $|D(\omega_i)\rangle = 1/\sqrt{2} (|V(\omega_i)\rangle + |H(\omega_i)\rangle)$ or $|A(\omega_i)\rangle = 1/\sqrt{2} (|V(\omega_i)\rangle - |H(\omega_i)\rangle)$. For $\varphi = -\pi/2$, the bi-photon output state becomes

$$|\Phi^{\varphi=-\pi/2}\rangle = 1/\sqrt{2} (|D(\omega_s) D(\omega_i)\rangle + e^{-i\pi/2} |A(\omega_s) A(\omega_i)\rangle) \qquad (2)$$

with no contribution from $|D(\omega_s) A(\omega_i)\rangle$ or $|A(\omega_s) D(\omega_i)\rangle$ states, as can be shown by replacing in (1) $|H(\omega_s)\rangle$, $|V(\omega_s)\rangle$, $|H(\omega_i)\rangle$ and $|V(\omega_i)\rangle$ by $|l_1(\omega_s)\rangle = 1/\sqrt{2} (D(\omega_s) + i A(\omega_s)\rangle)$, $|l_2(\omega_s)\rangle = 1/\sqrt{2} (i D(\omega_s) + |A(\omega_s)\rangle)$, $1/\sqrt{2} (|D(\omega_i)\rangle - |A(\omega_i)\rangle)$ and $1/\sqrt{2} (|D(\omega_i)\rangle + |A(\omega_i)\rangle)$, respectively. The $-\pi/2$ phase shift can be introduced as a path length mismatch ($\Delta L_s$) between the two arms of the interferometer traveled by the signal photons to ensure that both states always exit together through either port of the beam splitter (in that sense, the $-\pi/2$ phase shift compensates for the $\pi/2$ phase shift introduced by the beam splitter between the transmitted and reflected amplitudes). Alternatively, the $-\pi/2$ phase shift could also be introduced after recombination of idler beams at the PBS by using a quarter wave plate. In the latter case, idler photons would be circularly polarized and instead of linearly polarized states at +45°/-45°, one would measure instead clockwise/counter-clockwise output states in the conjugate basis.

The value of $\varphi = \varphi_p + \varphi_s + \varphi_i$ is determined by the phase delays accumulated by the pump, signal and idler along the two paths they respectively travel (from the PBS to the crystal for the pump, and from the crystal to either the 50-50 BS or the PBS for the signal and idler, respectively). For each of the three waves with wave vector $k_j$ ($j=p,s,i$), $\varphi_j$ is the delay $k_j \Delta L_j$ due to the arm length difference $\Delta L_j$. Recognizing that in free space $k_p = k_s + k_i$, and that $\Delta L_p = -\Delta L_i$ since pump and idler beams are counter-propagating, $\varphi$ is equal to $(-k_p + k_i) \Delta L_i + k_s \Delta L_s = k_s (-\Delta L_i + \Delta L_s)$. Instead of setting the value of $\varphi$ by stabilizing the path mismatch of the interferometers for signal ($\Delta L_s$) and idler ($\Delta L_i$) separately, one can stabilize the path difference between them ($-\Delta L_i + \Delta L_s$). Noting that the pump travels both idler ($\Delta L_i$) and signal ($\Delta L_s$) interferometers since dichroic mirrors used to extract the signal beams at 810 nm also have a residual reflectivity at the pump wavelength, it appears that $\varphi$ can be set by locking the Mach-Zehnder (MZ) interferometer traveled by the pump light between the PBS and the 50-50 BS where signal beams interfere (the path mismatch of this MZ interferometer is $-\Delta L_i + \Delta L_s$). This provides a convenient way to stabilize the phase against environmental disturbances, as illustrated in Fig. 2.

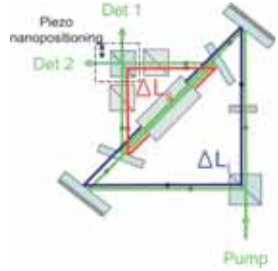

Fig.2. (Colors online). Principle of active stabilization of the single-crystal source. The output phase factor $\varphi = k_s (-\Delta L_i + \Delta L_s)$ can be set by locking the path mismatch $(-\Delta L_i + \Delta L_s)$ of the Mach-Zehnder interferometer (MZI) traveled by pump light between the polarizing beam splitter where it is split until the non polarizing beam splitter after which pump interferences can be monitored with photodiodes Det 1 and Det 2. For a preliminary alignment of the interferometer, we used a broadband source at 1550 nm and we looked at the output of the MZI with a spectrometer. Interferences modulate the continuous spectrum of the broadband source with oscillations, which get broader as the MZI gets balanced. The nearly equal-arm-length point is reached by means of a piezo nano-positioning actuator mounted on the interfering beam-splitter. The piezo-actuator is then used to tune the output phase factor to $-\pi/2$ for which maximum correlations are observed.

Apart from the conditions of spectral and spatial indistinguishability, which are fulfilled by the bidirectional pumping of the single-crystal and the coupling to single-mode fiber, good quality entanglement also requires temporal indistinguishability. Accordingly, the path-length difference between the two signal fields ($\Delta L_s$) should be the same as that between the two idler fields ($\Delta L_i$) within the coherence length of the photons (a few millimeters in our case). Setting the phase $\varphi = k_s(-\Delta L_i + \Delta L_s)$ to $-\pi/2$ leads to a path mismatch $(-\Delta L_i + \Delta L_s)$ of about 200 nm, a value which lies within the coherence length of the photons, such that the condition of temporal indistinguishability is fulfilled.

The avalanche photo-detectors (APD) used are four Si-based APD (Perkin Elmer SPCM-AQR-14) on the signal side (810 nm), having a quantum efficiency $\eta_s = 60\%$, and a home-built InGaAs-APD (Epitaxx) module on the idler side (1550 nm), having $\eta_i = 18\%$ quantum efficiency. The latter is gated upon detection at 810 nm with 2.5 ns-long pulses.

The source can be readily used for quantum key distribution, with the path-entanglement configuration at 810 nm achieving passive choice of basis in a more compact way than would be realized with polarization entanglement, for which signal beams would first need to be recombined at a PBS (after rotation of the polarization in one of the arms) before being split again at a 50-50 BS implementing the passive choice of basis.

## 3. Performance

Fig. 3 illustrates the quantum correlations shared by the entangled pairs after transmission over 100 m of single-mode fiber (SMF). Curves were obtained at a pump power $P = 1.2$ mW and single-count rate at 810 nm, $R_s = 0.3 \times 10^6$ s$^{-1}$. The coincidence rate reached a value $R_c$ above $1.1 \times 10^4$ counts/s for each of the four beam splitter output ports at 810 nm, yielding a conditional detection probability $R_c / R_s$ of about 3%. The rate of accidental counts, $R_a$, which we measured by random triggering of the InGaAs-APD, was about $0.5 \times 10^3$ s$^{-1}$, leading to a raw visibility $V_V = (R_c - R_a)/(R_c + R_a)$ of at least 91% in each basis. Taking into account the quantum efficiency $\eta_s = 60\%$ and $\eta_i = 18\%$ of the two APD, we estimate for the source a spectral brightness after coupling to the SMF $R_f \approx 4 \times R_c /(\eta_s \eta_i) \times 1/P \times 1/(k\,\Delta\lambda_i) \approx 3 \times 10^6$ s$^{-1}$ THz$^{-1}$ mW$^{-1}$, where $k = 0.125$ THz nm$^{-1}$ is the conversion factor between nm and THz, and where the factor of 4 arises from the four output ports at 810 nm. The quantum bit error rate (QBER), given by the averaged ratio $R_a / R_c$ over all four bases, amounts to about 5%. Using a narrower detection time window ($< 1$ ns) would reduce the rate of accidental counts and improve the visibility significantly.

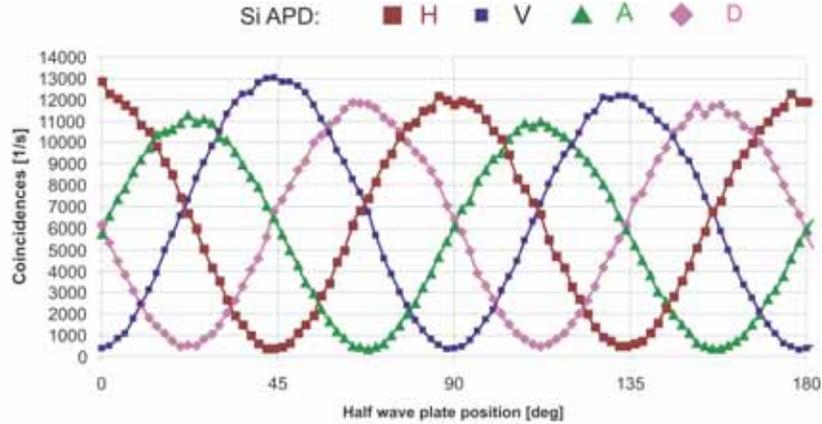

Fig.3. (Colors online). Coincidence rate after 100 m of fiber as a function of idler polarization for each of the four output states detected at 810 nm by the Si avalanche photo diodes (denoted H, V, D and A).

## 4. Conclusions

We have demonstrated a compact, narrowband, fiber-coupled source of path-polarization entangled photons at 810 and 1550 nm generated by bidirectional pumping of a 50 mm long PPLN:MgO crystal. The source has a spectral brightness of $3 \times 10^6$ $s^{-1}THz^{-1}mW^{-1}$ with two-photon interference visibility above 90%. With its small number of components, the source is ideal for table-top experiments. The set-up is compatible with applications requiring cw as well as pulsed operation. With 0.8 nm bandwidth at the telecommunication wavelength, the source can also be bridged over existing optical networks operating with 100GHz (0.8 nm) WDM environment. Moreover, the bandwidth is narrow enough so that real time polarization control of each flying qubit could be implemented with negligible contribution to the quantum bit error rate (QBER). By inserting a waveguide in a cavity with the same configuration, the bandwidth of the emitted pairs could be narrowed down to few hundreds of MHz with a coincidence rate still sufficiently high for enabling long distance teleportation schemes in which entanglement is stored in trapped-atom quantum memories, or to entangle photons from two independent sources.

**Acknowledgments**

We would like to thank Gunnar Björk for his careful reading of the manuscript. We also acknowledge support from HC Photonics for the periodically poled crystals made of LN:MgO. This work is supported by the European Commission through the integrated project SECOQC (Contract No IST-2003-506812) and QAP (Contract No IST-2005-015848), by the Swedish Foundation for Strategic Research (SSF) and the Swedish Research Council (VR), and by an equipment grant from Knut and Alice Wallenberg Foundation.